# Thulium- and Holmium-Doped Silica Fibres for 2 µm High-Power and Highly-Efficient Lasers

Richard Švejkar[a,*], Pavel Peterka[a,*], Bára Švejkarová[a,b], Michal Kamrádek[a], Petr Vařák[a,c], Martin Grábner[a], Jan Aubrecht[a], Ondřej Podrazký[a], Ivo Bartoň[a], Ivan Kašík[a] and Pavel Honzátko[a]

[a] *Institute of Photonics and Electronics of the Czech Academy of Sciences, Chaberská 57, Prague, 182 00, Czech Republic*
[b] *Faculty of Nuclear Sciences and Physical Engineering, Czech Technical University in Prague, Břehová 7, Prague, 115 19, Czech Republic*
[c] *Department of Inorganic Chemistry, University of Chemistry and Technology, Technická 5, Prague, 166 28, Czech Republic*

ABSTRACT

Silica optical fibres, originally developed as a low-loss transmission medium for telecommunications, have evolved into a versatile platform for compact and efficient high-power lasers. This review discusses recent progress in thulium- and holmium-doped active fibres emitting within the 2 µm spectral band, with an emphasis on material and waveguide approaches that address power scaling through thermal-load mitigation and enhanced laser efficiency. In particular, the temperature dependence of absorption and emission cross-sections are reviewed, and their implications for spectral gain, amplified spontaneous emission and high-power laser operation are discussed. Structured-core thulium fibres are then considered as a route to combining the high local $Tm^{3+}$ concentration required for efficient two-for-one cross-relaxation with a reduced average dopant concentration, lower longitudinal heat load, and a pedestal-free waveguide design. Finally, recent results on holmium-doped silica fibres prepared by nanoparticle-assisted modified chemical vapour deposition are summarised. Crucially, the Al/Ho ratio is identified as a key compositional parameter governing laser efficiency, while a high absolute alumina content supports a long fluorescence lifetime and enables high $Ho^{3+}$ concentrations without excessive clustering. These advances have enabled core-pumped holmium fibre lasers with slope efficiencies reaching 84 % and output powers approaching 100 W. Collectively, these developments illustrate how control of spectroscopy, composition, and transverse dopant distribution can extend the performance of silica fibre lasers across the 1.7-2.2 µm spectral region.

## 1. Introduction

The principle of optical-fibre guidance predates modern telecommunications and fibre technology by more than a century. In 1841, Jean-Daniel Colladon demonstrated that light could be guided along a curved water jet by total internal reflection [1]. At approximately the same time, Jacques Babinet extended the concept to light waveguide in bent glass rods [2]. A significant step towards practical glass-fibre guidance came in 1954, when van Heel demonstrated that coated glass fibres with a material of lower refractive index improved optical confinement by maintaining total internal reflection [3]. The invention of the laser in 1960 [4] then transformed the prospects for optical communication by providing a coherent optical source. Concurrently, Elias Snitzer proposed a fibre laser [5] and shortly afterwards realised this idea in a few-mode Nd-doped barium crown glass [6]. Shortly afterwards, group of Czech-born researcher Erich Spitz and Jean-Claude Simon at CSF in France prepared the very first single-mode fibre and investigated its potential for guiding coherent light and optical communication [7, 8].

Now, the main question was no longer whether light could be guided in a fibre, but whether glass could be made sufficiently transparent. An important observation came from glass science in 1965; Fiona Steele and Ronald Douglas quantified the optical absorption associated with iron in silicate and borate glasses [9], providing direct evidence that transition-metal impurities could contribute strongly to optical losses. Charles K. Kao and George A. Hockham at Standard Telecommunication Laboratories brought this materials perspective together with a detailed analysis of dielectric-fibre waveguides in their landmark 1966 paper [10]. They showed that bending, radiation and waveguide imperfection losses did not fundamentally prevent long-distance optical transmission. Instead, they argued that the high attenuation measured in fibres at that time was mainly caused by removable material impurities [9]. Their analysis indicated that losses as low as approximately $20\,dB{\cdot}km^{-1}$ should be achievable, establishing a realistic target for optical communications and representing the main significance of the 1966 paper [10].

Kao's research group subsequently developed increasingly sensitive spectrophotometric methods for measuring very small optical losses in bulk glasses [11–13], supporting the identification of fused silica as a material of sufficiently high purity for optical communication [14]. In 1970, the Corning team led by Robert D. Maurer translated this concept into low-loss silica fibre, reaching the attenuation level required for practical optical communication [14, 15]. The subsequent refinement of glass purification, vapour-deposition methods, preform fabrication and fibre-drawing

svejkar@ufe.cz (R. Švejkar); peterka@ufe.cz (P. Peterka)
ORCID(s): 0000-0002-2421-2177 (R. Švejkar); 0000-0002-6341-7589 (P. Peterka); 0009-0006-3604-2319 (B. Švejkarová); 0000-0003-0402-1872 (M. Kamrádek); 0000-0001-8026-0686 (P. Vařák); 0000-0003-1399-094X (M. Grábner); 0000-0002-3395-4375 (J. Aubrecht); 0000-0003-4255-9366 (O. Podrazký); 0000-0002-5100-5931 (I. Bartoň); 0000-0002-0887-9756 (I. Kašík); 0000-0003-4649-4790 (P. Honzátko)

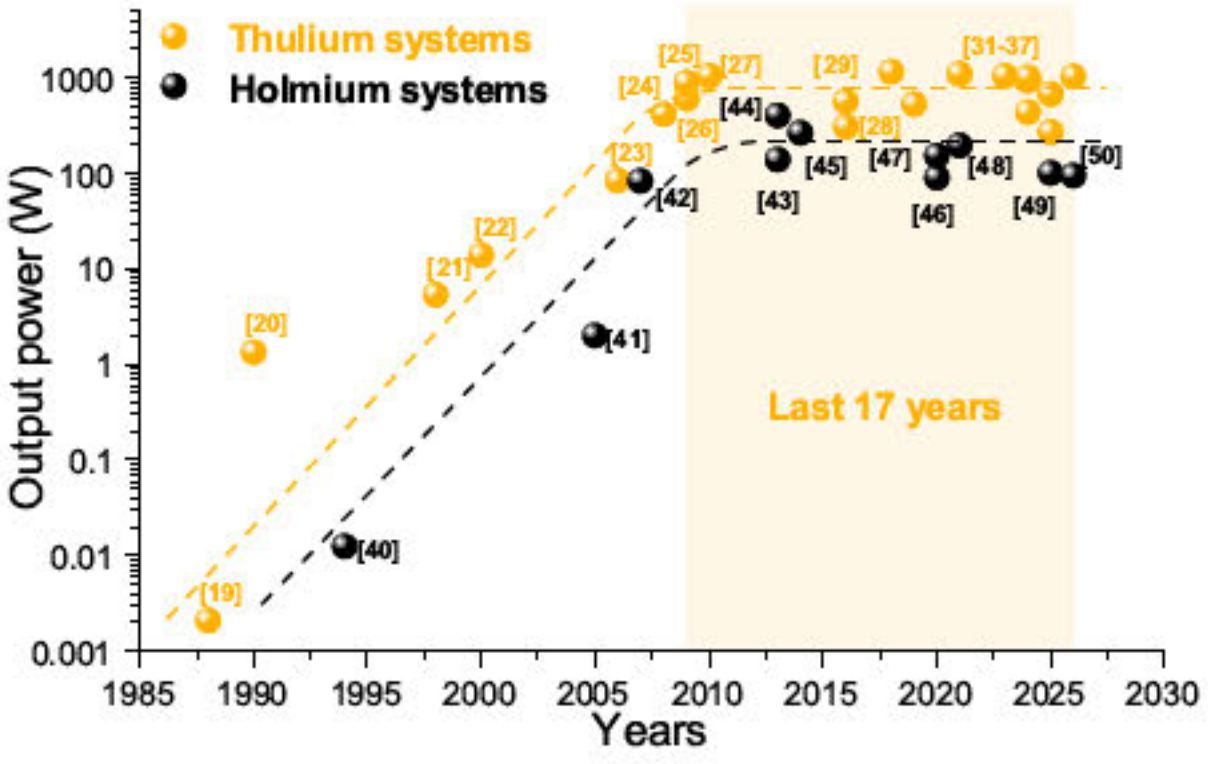

**Figure 1:** Power evolution of thulium-based (yellow) and holmium-based (black) fibre lasers and amplifiers over approximately the past 40 years [19–50].

technologies established the technological platform for modern optical fibres. In parallel, research on rare-earth-doped fibre lasers, which had begun with Nd-doped fibres in the 1960s [6, 16], benefited directly from these advances. David Payne, Simon Poole, Martin Fermann, Anne Tropper and other researchers at the University of Southampton demonstrated [17] that controlled concentrations of rare-earth ions could be incorporated into low-loss silica fibres using an extended modified chemical vapour deposition (MCVD) process, enabling the development of active fibres, optical amplifiers and fibre lasers [17, 18].

In this paper, we review rare-earth thulium- and holmium-doped active fibres operating at emission wavelengths around 2 µm, whose development dates back to the late 1980s and the 1990s, respectively, and builds on the advances in rare-earth-doped silica fibres outlined above. Figure 1 summarises the evolution of output power over approximately the past 40 years [19–50]. Several examples are provided from our research how temperature-dependent absorption cross-sections can influence laser operation, and how structured-core design and optimisation of the optical materials can mitigate high thermal loads and improve laser efficiency.

### 1.1. Thulium-doped fibre lasers

The trivalent thulium ion, $Tm^{3+}$, has a complex energy-level structure and supports laser emission in the wavelength range of 1.7-2.2 µm. The main applications of these lasers include medicine [51], materials processing (particularly polymers) [52], spectroscopy [53], nonlinear frequency conversion towards longer wavelengths in the mid-infrared spectral region [54], LIDAR [55], and defence applications [24, 56].

The first continuous-wave operation of a single-mode Tm-doped fibre laser was demonstrated in 1988 [19], establishing the feasibility of using thulium-doped glass fibres as laser gain media in the 2 µm spectral region. Two years later, the continuous-wave output power was increased to the 1 W level [20]. In this early experiment, pump radiation from a Nd:YAG laser at 1064 nm was coupled directly into the doped fibre core, and the limited pump power that could be launched into the small core restricted further power scaling.

A major step towards high-power operation therefore came with the introduction of double-clad fibre geometries and cladding pumping [57]. In 1998, cladding-pumped operation of a Tm-doped silica fibre laser was demonstrated, allowing lower-brightness laser-diode pump sources to be used [21]. By 2000, efficient cladding-pumped operation with 14 W of single-mode output power had been achieved [22]. The transition from core pumping to cladding pumping was thus one of the key developments that provided a practical route towards substantial power scaling.

During the following decade, pumping in the 790 nm wavelength band became established as the principal approach for power-scalable Tm-doped silica fibre lasers [27–31]. Moreover, the importance of high alumina co-doping to reduce $Tm^{3+}$ clustering and quenching was recognised [58]. The combination of double-clad fibre technology, high-power 790 nm laser diodes and efficient energy-transfer processes in highly Tm-doped silica formed the architecture that remains dominant in high-power Tm-doped fibre lasers today [22, 23].

Although pumping at 790 nm excites the $^3H_4$ manifold rather than directly the upper laser level $^3F_4$, efficient energy transfer can occur in sufficiently highly doped fibres. Owing to the well-known two-for-one cross-relaxation process, illustrated in Figure 2, can generate up to two ions in the upper laser level for each absorbed pump photon, resulting in a theoretical maximum quantum efficiency of 2 [29, 59]. For pump and laser wavelengths of 790 nm and 1940 nm, respectively, the corresponding theoretical maximum optical efficiency is approximately 81 %. Efficient utilisation of this process requires a relatively high concentration of $Tm^{3+}$ (> 3.5 wt. %), which promotes efficient energy transfer between neighbouring ions and hence efficient cross-relaxation ($^3H_4$,$^3H_6$ → $^3F_4$,$^3F_4$) [60]. This mechanism has enabled the development Tm-doped fibre lasers with output powers ranging from several hundred watts to the kilowatt level, both in oscillator configurations [24, 29] and in master-oscillator power-amplifier (MOPA) architectures [33–35].

Several alternative pumping schemes have also been investigated. Tm-doped active fibres can be resonantly core-pumped in the 1.5-1.9 µm region, where the upper laser level $^3F_4$ is excited directly through the $^3H_6$ → $^3F_4$ transition. Core pumping commonly uses Er-doped fibre lasers operating around 1.55 µm and can enable laser generation towards the shorter-wavelength side of the thulium emission band [61, 62], although power scaling is constrained by the available Er-doped fibre-laser pump power. Direct cladding pumping at 1710 nm also excites the upper laser level with a low quantum defect and potentially high efficiency [63], but is currently limited by the relatively low brightness and output power of available laser diodes, approximately 80 W [37]. More recently, wavelength-shifted Raman lasers operating around 1650 nm have enabled optical-to-optical efficiencies of approximately 80 % and output powers exceeding 1.2 kW [64]. This approach, however, increases system complexity and the thermal burden associated with pump generation is transferred to the Raman laser pump stages. Consequently, despite the high efficiency of the 1650 nm-pumped thulium stage itself, the overall system efficiency remains lower than for direct pumping at 790 nm. For this reason, direct cladding pumping with high-power laser diodes near 790 nm remains the most widely used route to kilowatt-class operation [33]. Tm-doped fibres can also generate laser emission at shorter wavelength around 1.7 µm, although strong ground-state reabsorption and gain competition with longer-wavelength emission make efficient operation more challenging. Recent advances in pumping configurations, fibre-length optimisation and suppression of amplified spontaneous emission, e.g. using depressed fibres [65], have enabled significant progress in this spectral region [66]. The single-frequency amplification exceeding 30 W over the 1730–1760 nm range has recently been demonstrated[67].

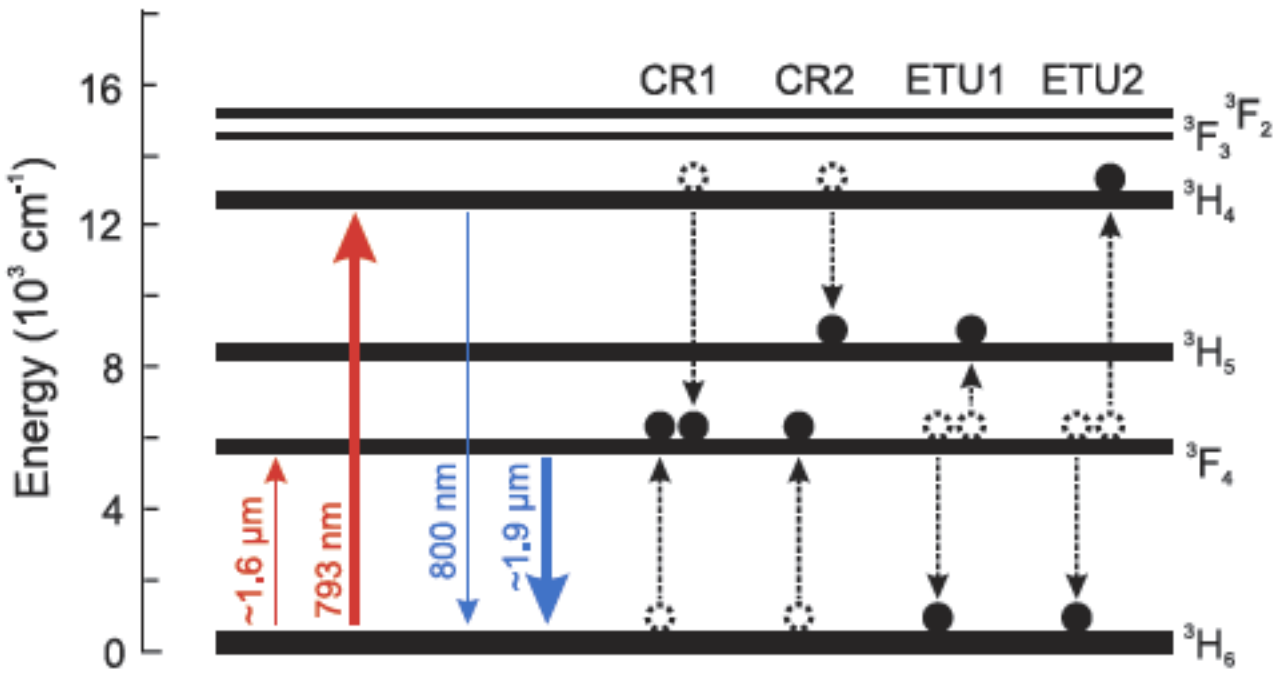

**Figure 2:** Energy levels and transfer processes in $Tm^{3+}$.

Figure 2 summarises the principal energy-transfer processes in Tm-doped active media. Cross-relaxation, particularly CR1, is beneficial because it is responsible for the two-for-one process [59]. The strength of CR1 depends primarily on the separation between neighbouring $Tm^{3+}$ ions and therefore on the local thulium concentration [59]. CR2 is weaker because it requires multiphonon assistance. In contrast, the energy-transfer upconversion processes ETU1 and ETU2 depopulate the $^3F_4$ manifold and therefore reduce laser efficiency. Although both ETU processes also require multiphonon assistance and are weaker than CR1, they can still influence the effective efficiency of the cross-relaxation process [59].

The subsequent development of high-power Tm-doped systems was characterised by rapid scaling of both oscillators and amplifiers. A key challenge for large-mode-area (LMA) fibres is that the high alumina and thulium concentrations required for efficient 790 nm pumping increase the refractive-index contrast, whereas LMA operation requires a low numerical aperture (NA). This conflict was addressed by the so-called pedestal design [68], in which the doped core is surrounded by a region of elevated refractive index, reducing the effective core-to-surrounding index contrast and thereby providing the low NA required for LMA operation. Although pedestal-based fibres are now widely used in high-power thulium-doped lasers and amplifiers, they introduce additional challenges, including splice sensitivity and possible excitation of pedestal modes [32, 34]. Pedestal-free approaches have therefore been developed, including structured-core designs [69, 70] and optimisation of glass composition [71–73]. Recent developments by Coherent have demonstrated single-mode operation of a pedestal-free Tm-doped fibre with a 25 µm core and efficiency exceeding 50 % [69]. Similar pedestal-free fibres have also been introduced commercially by Coractive. These developments highlight an alternative approach to high-power fibre design, in which thulium concentration and glass composition are optimised to balance cross-relaxation efficiency and waveguide properties without the additional complexity of a pedestal structure.

By 2009, single-frequency Tm-doped fibre amplifiers, utilising LMA fibres, had reached output powers above 600 W while maintaining single-mode operation and high beam quality [26]. Kilowatt-class operation followed shortly afterwards [27]. Further advances in LMA fibres, fibre Bragg gratings, pump combiners, thermal management and all-fibre integration subsequently enabled monolithic oscillators with output powers above 500 W [29]. In parallel, MOPA architectures became increasingly important because they allowed the seed wavelength and spectral properties to be defined independently of the subsequent power-amplifier stages. Tm-doped fibre amplifiers have reached and exceeded the kilowatt level while preserving selected seed characteristics and near-diffraction-limited beam quality [26, 32, 36, 38].

Recent development has shifted from demonstrating high-power feasibility towards simultaneously optimising output power, efficiency, spectral properties and system integration. Kilowatt-class MOPA systems, narrow-linewidth sources [36], widely tunable high-power systems [33] and monolithic all-fibre oscillators above the kilowatt level [39] have all been demonstrated. At the same time, increasing attention has been focused towards optical efficiency and thermal management [37, 70]. Conventional uniformly doped Tm fibres typically operate substantially below the theoretical efficiency limit [33, 35], as illustrated in Figure 3. Current approaches therefore include optimisation of thulium concentration and glass composition [60], spatial tailoring of the active-ion distribution, and structured-core designs [37, 74, 70] that aim to balance pump absorption, cross-relaxation, propagation loss and thermal loading [60, 70].

Despite these advances, the output power of Tm-doped fibre lasers and amplifiers has remained close to the kilowatt level for approximately 17 years, as shown in Figure 1. Further scaling is increasingly constrained by the substantial heat load generated in the active fibre. At high pump powers, effective heat removal becomes critical because excessive temperature can lead to coating damage, thermally induced modal effects and, ultimately, fibre failure. Reducing or redistributing the generated heat load is therefore essential for extending performance beyond present levels. Under

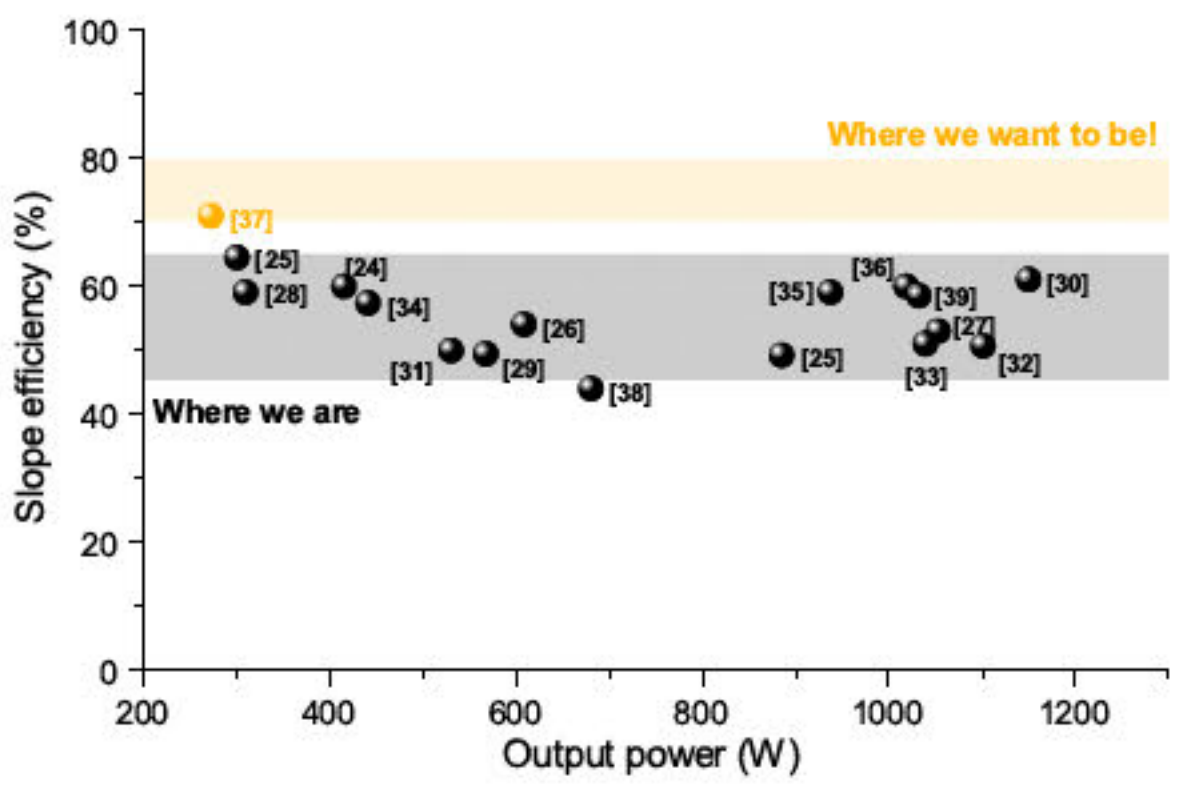

**Figure 3:** Slope efficiency evolution of thulium-based fibre lasers and amplifiers with increasing output power [24–39]. The yellow dot represents results for structured nested-ring fibre [37].

790 nm pumping, demonstrated efficiencies also remain substantially below the theoretical limit. The highest published output power from a single-mode, monolithic, all-fibre Tm-doped oscillator is currently 1032 W, with an efficiency of 58.5 % with respect to absorbed pump power [39]. The highest reported efficiency at high output power is 71 %, corresponding to 272 W of output power [37]. In a power-amplifier configuration, output powers of up to 1.1 kW have been demonstrated [30, 32], with an efficiency of 61 % reported in [30].

## 2. Thulium-doped fibre spectroscopy and temperature-dependent cross-sections

The fluorescence lifetimes of emissions from the $^3F_4$ and $^3H_4$ levels provide important insight into the local environment of the Tm ions, ion clustering, and the dynamics of energy-transfer and other non-radiative processes, all of which influence the suitability of the fibre for laser operation. Vařák et al. [72] recently demonstrated that longer lifetimes of the $^3F_4$ manifold indicate low rates of concentration quenching and multiphonon relaxation, whereas the $^3H_4$ emission is highly sensitive to the cross-relaxation process, which depopulates the $^3H_4$ level and shortens the corresponding lifetime. In principle, the combination of a long $^3F_4$ lifetime and a short $^3H_4$ lifetime indicates a potentially promising fibre for two-for-one laser operation. Moreover, measured fluorescence decay curves can be used to directly determine energy-transfer coefficients through rate-equation modelling, as demonstrated by Kamrádek et al. [73].

As discussed above, the high heat load in Tm-doped fibres can raise the active-core temperature to several hundred degrees Celsius under realistic high-power operating conditions [75]. Despite the well-known temperature dependence of rare-earth absorption and emission, corresponding cross-section data for Tm-doped silica over this temperature range have only recently become available. Švejkarová (Jiříčková) et al. measured the spectra associated with the $^3H_6 \rightarrow {}^3H_4$ and $^3H_6 \rightarrow {}^3F_4$ transitions, i.e. the manifolds involved in the two-for-one cross-relaxation process, over a temperature range from liquid-nitrogen temperature, 77 K (-196 °C), up to 575 K (300 °C) [76, 77]. Multiple Gaussian fits of representative spectra are summarised in Figure 4. The spectra and fitting parameters are available in the Supplementary Material and in [78]. It should be noted that the spectra in Figure 4 were obtained using a liquid nitrogen cryostat [76]. In the first publication [77], the cryogenic emission measurements were carried out with the fibre under test in an open bath of liquid nitrogen, which provided less stable temperature conditions. Although these issues were largely reduced in the later cryostat measurements [76], some aspects of temperature-dependent emission measurements of thulium fibres are not yet fully resolved. Possible contributions include temperature-dependent absorbed excitation power, dependence on excitation wavelength, changes in fluorescence lifetimes and cross-relaxation coefficients, and an unexpected abrupt change in the bend loss of G.652 fibre pigtails at low temperatures [79].

The importance of these data becomes evident when they are incorporated into laser modelling. Even under simplified assumptions, temperature-dependent spectra predict substantial and sometimes non-intuitive changes in laser behaviour. For example, for a constant fibre-core temperature and a fixed relative population of 5 % in the metastable $^3F_4$ manifold, the gain peak shifts from approximately 1940 nm at room temperature to approximately 2020 nm at 200 °C [80]. A first estimate of the corresponding changes in laser threshold and slope efficiency can be obtained using an analytical model based on the rate and propagation equations, including the two-for-one cross-relaxation process [81]. In the initial study, the slope efficiency evaluated at the fibre length corresponding to 13 dB pump absorption, i.e. approximately 95 % absorbed pump power, decreased by only about 2-4 % when the core temperature increased from 23 to 300° C [80, 82]. This analytical model does not, however, account for amplified spontaneous emission (ASE). In configurations operating near the edge of the emission spectrum, temperature-dependent spectroscopy can have a much stronger effect: self-heating shifts the gain spectrum towards longer wavelengths and can substantially increase ASE, potentially making it comparable to or larger than the intended signal and strongly suppressing laser emission at the target wavelength [82]. Current numerical studies also largely neglect the temperature dependence of other spectroscopic parameters, including fluorescence lifetimes and energy-transfer coefficients. Incorporating these dependencies should improve the predictive capability and optimisation of high-power thulium fibre lasers and amplifiers.

## 3. Core structuring in thulium-doped fibres

The high local $Tm^{3+}$ concentration required for efficient two-for-one cross-relaxation contributes to the substantial heat load generated in high-power Tm-doped fibres. A conventional step-index fibre with uniform rare-earth doping is therefore not necessarily optimal when thermal management and propagation loss become limiting. Spatial structuring of the active core region represents one possible strategy for mitigating thermal loading and propagation loss-related efficiency limitations in Tm-doped fibres [37]. Several approaches can be used to structure the active region, as schematically illustrated in Figure 5. They can be divided into one-, two- and three-dimensional modifications, where this designation follows the three cylindrical coordinates $r$, $\varphi$ and $z$ used in optical fibres. The one-dimensional (1D) approach preserves cylindrical symmetry and modifies the radial distribution of the rare-earth ions, for example through concentric or nested-ring structures [70]. In the two-dimensional (2D) approach, the active core is composed of multiple individual elements, typically rods, which are assembled during preform fabrication and subsequently drawn into the fibre [74, 83]. In some applications, the size of rods matters, for reducing diffusion effects, the rods should be as big as technology allows as illustrated in Figure 5, while for smooth refractive index profile the nanostructuring approach is preferable, as shown in Figure 5. This is so-called stack-and-draw technique [84]. The third approach extends structuring into the third dimension, in which the rare-earth concentration is varied along the longitudinal $z$-axis of the active fibre rather than in the radial direction [85, 86].

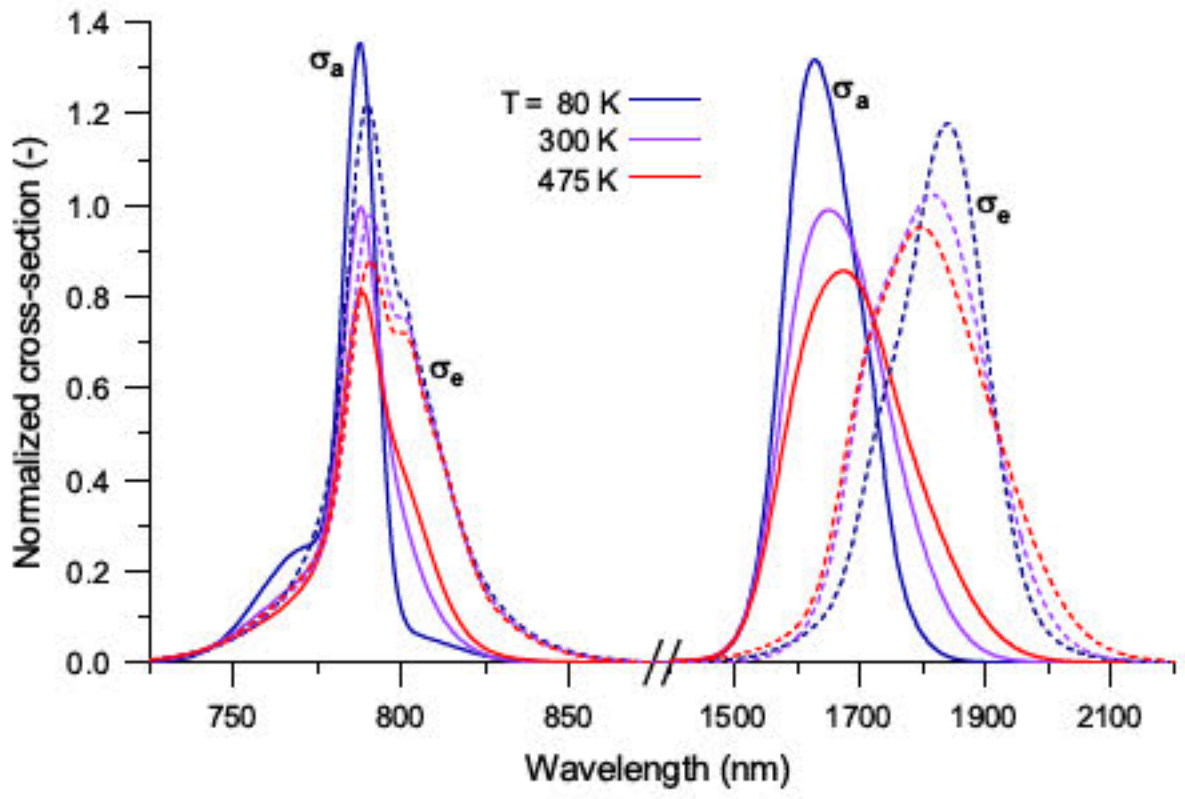

**Figure 4:** Absorption (solid lines) and emission (dashed lines) cross-section spectra of thulium at selected temperatures. The curves were calculated using the procedures described in [37, 77].

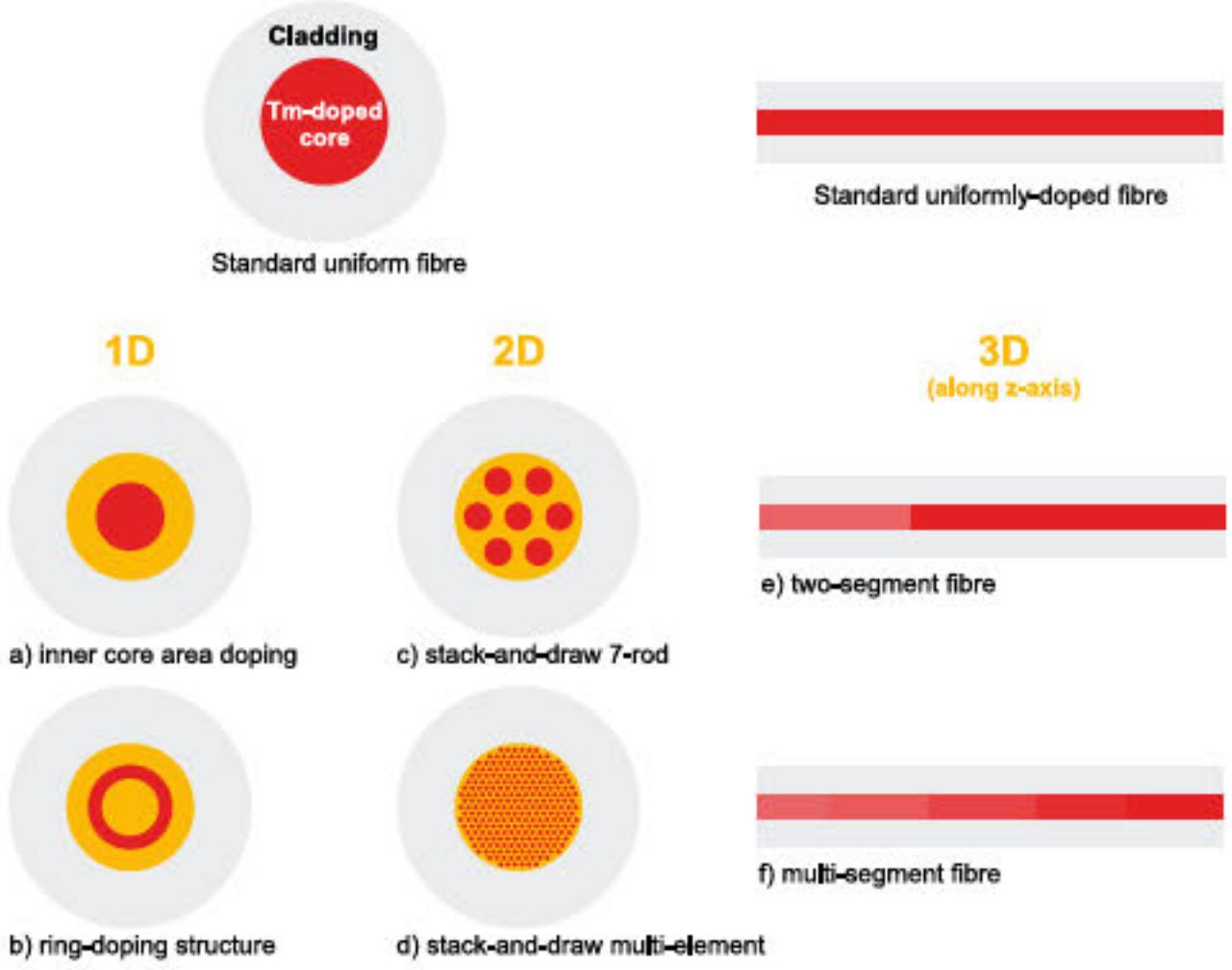

**Figure 5:** Various structured-core designs for Tm-doped fibres. The red regions represent the Tm-doped areas, the yellow regions the undoped parts of the core, and the grey region the cladding.

Rather than distributing the $Tm^{3+}$ uniformly in the core, the radial distribution of the active ions can be tailored to modify the local pump absorption and heat generation. The direct influence of such an approach has been demonstrated in Tm-doped 'nested-ring' fibres [37, 70], while further advantages of structured active regions have been investigated using more complex core geometries [74].

The radial structuring approach (1D) enabled an output power of 272 W with a slope efficiency of 71 % from a Tm-doped active fibre with core and cladding diameters of 8.2 µm and 135 µm, respectively. This remains the highest reported output power for a Tm-doped fibre with a core diameter below 15 µm and the highest published slope efficiency for a Tm-doped fibre laser operating above 50 W [37]. This performance was enabled by tailoring the doped region into a nested-ring structure, in which thulium ions were confined to a ring approximately 1 µm thick. This design preserves a high local $Tm^{3+}$ concentration, thereby promoting efficient two-for-one cross-relaxation, while reducing the overall amount of doped material and consequently the thermal load generated during laser operation. This result can therefore be regarded as a proof of concept demonstrating the potential benefits of structured-core designs for high-power operation around 2 µm.

An alternative to the radially symmetric modification of the dopant distribution is the multi-rod core concept, for example the seven-rod-core configuration shown in Figure 5. In this design, the core consists of seven Tm-doped rods embedded within an otherwise undoped core region. In [74], it was demonstrated for the first time that a structured-core approach can support Tm-doped fibre lasers with relatively high slope efficiencies of up to 55 % and output powers of 25 W. This proof of concept study established the potential of the multi-rod architecture and paved the way for further research showing slope efficiency of 72 %, at an output of 55 W at 1907 nm [87].

These stack-and-draw multi-element preforms are fabricated using the conventional MCVD method through nanoparticle-solution doping and with an optimised Al/Tm ratio to maintain a sufficiently high local thulium concentration for efficient cross-relaxation energy transfers. The preform core itself was assembled from seven Tm-doped canes, which formed the active region of the fibre core, with a diameter of approximately 10 µm. Note that issue of dopant diffusion should be carefully assessed, because in the case of too small rod elements ($< 1$ µm) the diffusion may reduce the local $Tm^{3+}$ concentration that does not any more fulfil the requirements for the efficient two-for-one process [88]. Further details of the fibre fabrication process are reported in [74].

Unlike radial or multi-rod structuring, a continuous longitudinal variation of the dopant concentration is not readily realised during conventional fibre drawing. This approach may become feasible in the future if 3D preform printing technologies advance sufficiently [89]. An equivalent longitudinal concentration profile can, however, be formed by splicing together active-fibre sections with different $Tm^{3+}$ concentrations [85]. Such an arrangement enables the axial distribution of pump absorption and heat generation to be tailored, particularly near the beginning of the active fibre, where the pump power are typically highest.

This approach was investigated experimentally for the first time in 2025 in [85] as a joint effort between the Institute of Photonics and Electronics of the Czech Academy of Sciences and the Optoelectronics Research Centre at the University of Southampton. The main idea was to mitigate thermal limitations by introducing a longitudinally varying $Tm^{3+}$ concentration along the active fibre. In this configuration, the pump radiation first interacts with a lower-doped section of the fibre and then progressively encounters sections with higher thulium concentration. This redistributed pump and consequently reduced the local temperature maximum that normally occurs close to the pump-input end [85].

Several important outcomes were demonstrated experimentally. The additional splices within laser cavity did not significantly affect the laser performance; with the segmented fibre reaching a slope efficiency of 65 %, compared with 65.3 % for the uniformly doped reference fibre [85]. At the same time, thermal-camera measurements showed a more uniform temperature distribution and a significantly lower maximum temperature. As a result, the segmented fibre could withstand higher pump powers, whereas the uniformly doped fibre suffered thermal failure above approximately 40 W of pump power. The fibre laser with segmented active fibre composed of four segments of different concentration ultimately produced 54 W with a slope efficiency exceeding 63 % [85]. This is competitive result with other published values for Tm-lasers [35, 44, 59].

These results confirmed that longitudinal modification of the $Tm^{3+}$ concentration can provide an additional degree of freedom for thermal management in high-power Tm-doped fibre lasers. By distributing pump absorption more uniformly along the active fibre, local heat-load extremes can be reduced while maintaining a sufficiently high local $Tm^{3+}$ concentration to preserve efficient two-for-one cross-relaxation. The demonstrated segmented-fibre approach therefore represents a practical approximation to an ideally continuous longitudinal concentration gradient, which remains difficult to realise directly with current fibre-fabrication technology [85].

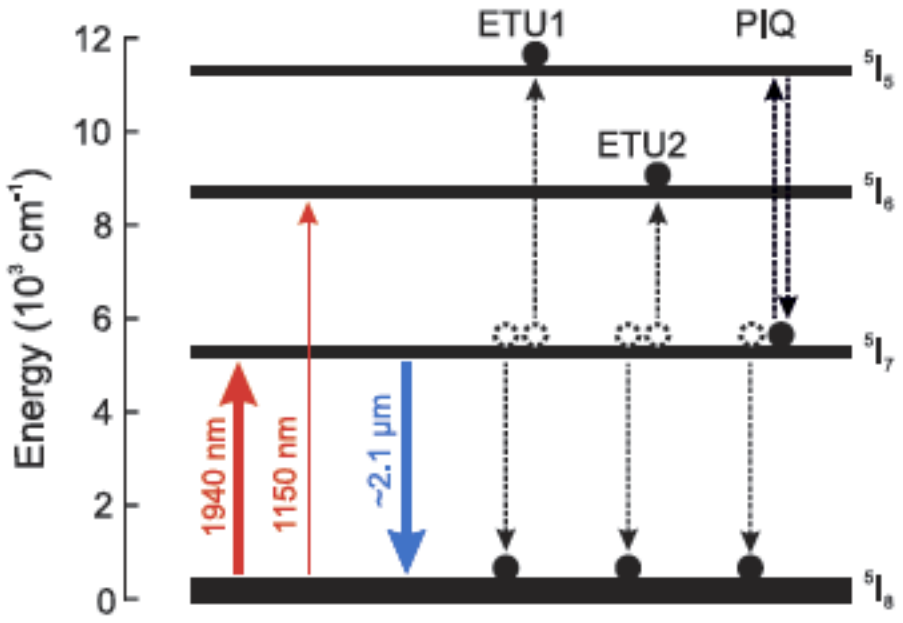

**Figure 6:** Energy levels of $Ho^{3+}$ and the principal energy-transfer processes.

## 4. Holmium-doped fibre lasers

Ho-doped fibre lasers and amplifiers operate mainly in the 2.1-2.2 μm region and address a range of applications similar to those of the Tm-doped systems discussed above [43, 90]. A key spectroscopic feature of $Ho^{3+}$ in silica is the broad absorption band centred near 1950 nm, corresponding to the $^5I_8 \rightarrow {}^5I_7$ transition, as illustrated in Figure 6. This band is well matched to Tm-doped fibre lasers and therefore enables efficient resonant pumping [46, 91]. The corresponding absorption and emission cross-sections are shown in Figure 7. The development of high-power Ho-doped fibre lasers (HDFLs) has consequently remained closely linked to progress in Tm-doped pump sources.

Although the emission bands of Tm- and Ho-doped fibre lasers overlap around 2.0–2.1 μm, their practical operating regimes differ. Tm-doped systems are particularly suitable for high-power operation around 2.0–2.1 μm, benefiting from direct 790 nm diode pumping, with 937 W and 59 % slope efficiency demonstrated at 2036 nm [35]. Ho-doped fibres can operate down to 2025 nm [90], although ground-state reabsorption becomes increasingly important towards shorter wavelengths. Due to the additional complexity associated with using Tm-doped fibre lasers as pump sources, Ho-doped fibres are particularly advantageous at wavelengths beyond approximately 2.1 μm, where Tm-doped systems become less efficient.

The first continuous-wave Ho-doped silica fibre laser, demonstrated in the first half of the 1990s, used a Tm-sensitised Ho-doped fibre [40]. In this approach, Tm ions absorbed the pump radiation and transferred the excitation energy to $Ho^{3+}$, providing a practical pumping route before suitable high-power sources for direct resonant excitation of the holmium absorption band were available.

In 2005, an efficient and wavelength-tunable Ho-doped silica fibre laser was demonstrated, providing a tuning range of 160 nm and a maximum output power of 4.3 W with a slope efficiency of approximately 45 % [41]. This was followed in 2007 by a substantial increase in output power, when 83 W was generated from a Tm, Ho co-doped double-clad silica fibre laser using 793 nm diode pumping [42]. The laser achieved a slope efficiency of 42 % with respect to launched pump power while maintaining good beam quality. These results demonstrated the potential of Ho-based silica fibre systems for efficient, high-power laser generation around 2.1 µm under direct laser diode pumping.

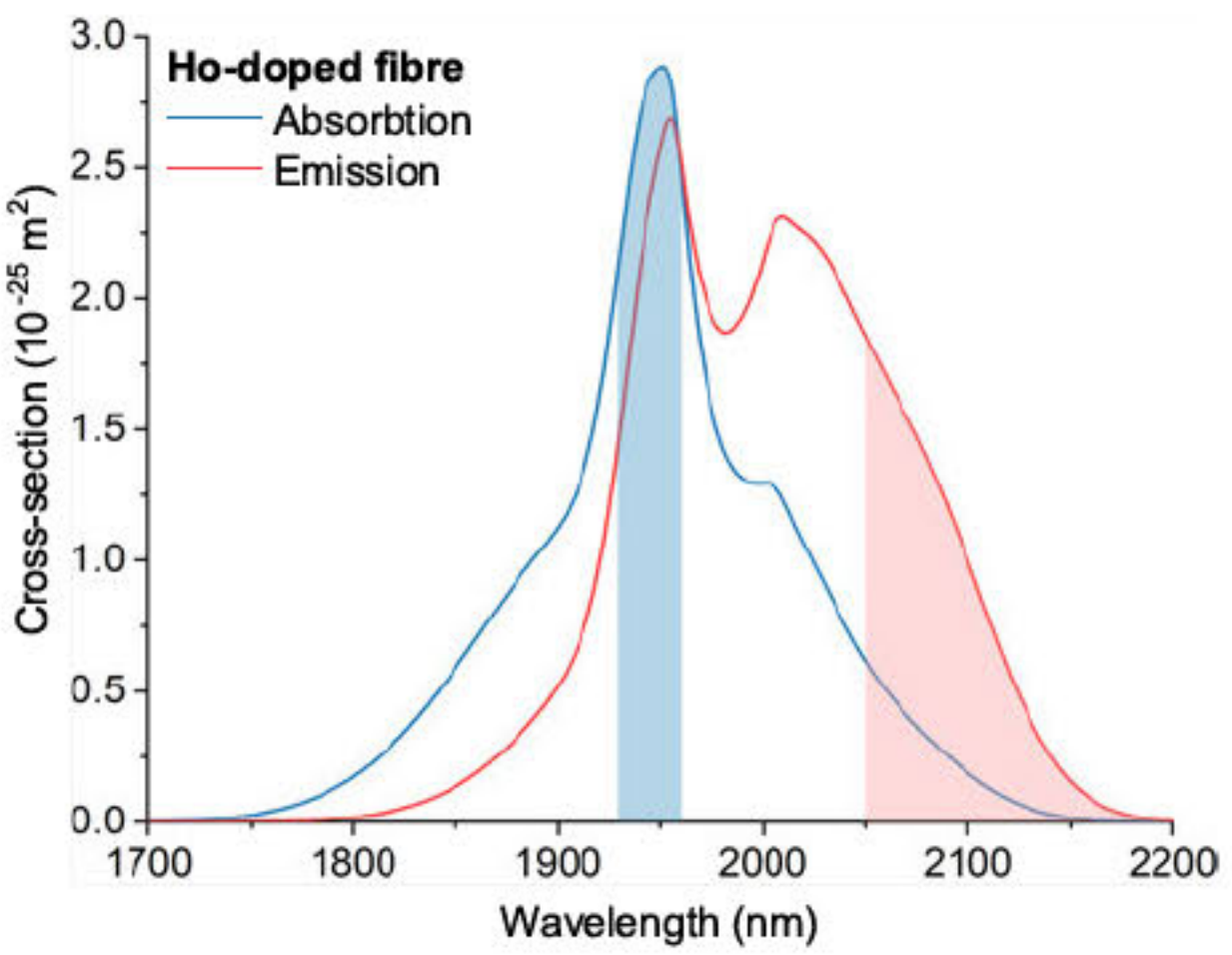

**Figure 7:** Absorption and emission cross-sections of Ho-doped active fibre with 3850 mol ppm of $Ho^{3+}$ [91].

A further step towards power scaling was the development of all-glass cladding architectures suitable for high pump powers around 1.9-1.95 µm [43, 92]. In triple-clad implementations, an additional glass cladding isolates the outer polymer coating from direct exposure to the pump radiation, which is strongly absorbed by conventional polymer coatings and could otherwise cause excessive heating or damage. In 2013, an all-glass cladding-pumped Ho-doped silica fibre laser resonantly pumped by a Tm-doped fibre laser near 1.95 µm produced up to 140 W, with a maximum slope efficiency of 59 % with respect to launched pump power [43]. The same resonant-pumping concept was subsequently extended to amplifier configurations, and in 2014 an LMA Ho-doped fibre amplifier produced 265 W at approximately 2.11 µm [45].

Subsequent work addressed both power scaling and application-specific performance. High-power Ho-doped fibre lasers for long-range transmission were investigated experimentally and numerically in 2020 [46], reflecting interest in the atmospheric transmission properties of radiation near 2 µm. In parallel, Tm, Ho co-doped fibres continued to provide an alternative route towards high-power operation under 790 nm pumping through thulium. A monolithic single-oscillator system produced 150 W in 2020 [47], followed by a diffraction-limited output power of 195 W with a slope efficiency of 44.8 % at 2.09 µm in 2021 [48]. These results show that co-doped fibre architectures can be competitive for obtaining high-power and high-beam quality laser radiation, although, with efficiencies significantly below theoretical limit.

Recent research have also addressed practical limitations associated with pump delivery, fibre integration and the overall efficiency of high-power HDFL architectures. In 2025, a high-power 7×1 triple-clad fibre combiner specifically developed for resonant cladding pumping of Ho-doped fibre lasers at 1.94 µm was demonstrated [49]. The component was subsequently integrated into an all-fibre Ho-doped laser producing 101 W at 2.12 µm with a slope efficiency of 61 % with respect to launched pump power. This architecture illustrates the continuing effort towards fully integrated, all-fibre and power-scalable Ho-doped laser systems [49].

At the same time, core-pumped HDFLs have attracted renewed attention when sufficiently high-brightness pump sources are available. In 2026, an all-fibre core-pumped Ho-doped fibre laser generated 95 W with a slope efficiency of 84 % was demonstrated [50]. The key technology behind this result is the nanoparticle doping method which is in detail discussed in the following section 5. This result illustrates an alternative development path in which the lower pump-power handling capability of core pumping can be compensated by high pump brightness, strong pump-signal overlap and a low quantum defect, favouring high efficiency rather than maximum output power as the principal advantage.

The high efficiency of resonantly pumped HDFLs originates primarily from the small separation between the pump and laser wavelengths. Pumping at 1940-1950 nm directly excites the upper laser level $^5I_7$, while laser emission typically occurs in the 2.1-2.2 µm range. The corresponding quantum defect is below approximately 8 % [93], limiting the fraction of absorbed pump power converted into heat and enabling high optical efficiencies. The resulting low heat load also facilitates thermal management, including in core-pumped systems [50, 94]. Energy-transfer upconversion can nevertheless reduce performance. As illustrated in Figure 6, ETU1 and ETU2 involve the interaction of two excited Ho ions, with one ion promoted to a higher-lying level ($^5I_5$ for ETU1 or $^5I_6$ for ETU2) while the other returns to the $^5I_8$ ground manifold [95]. Both processes therefore depopulate the $^5I_7$ upper laser level [91, 95]. Cross-relaxation and excited-state absorption are considered negligible under resonant pumping in the systems considered here [95].

Core- and cladding-pumped Ho-doped fibre lasers therefore present different scaling trade-offs. Core pumping provides strong spatial overlap and can operate with active-fibre lengths below 1 m, reducing accumulated signal propagation losses and enabling very high efficiencies [50]. Its principal limitation is the pump power that can be coupled into the single-mode core. Cladding pumping can accept substantially higher pump powers and has consequently enabled the highest output powers, including 407 W with a slope efficiency of 40 % using a kilowatt-class Tm-doped fibre laser as the pump source [44]. However, cladding-pumped systems typically require several metres of active fibre, and in some cases more than 10 m, so propagation losses at 2 µm can accumulate and contribute to slope efficiencies of only about 40-60 % [43, 44, 49, 60]. This creates a slope-efficiency difference exceeding 25 % between representative core- and cladding-pumped systems, as shown in Figure 8. As an intermediate architecture, pedestal pumping has been proposed for Ho-doped fibre lasers [96], analogous to the concept previously investigated for Tm-doped systems [97]; here, the pump is guided in a relatively small inner cladding forming the pedestal waveguide. Further scaling of resonantly pumped Ho-doped fibre lasers nevertheless remains closely linked to the continued development of high-power and high-brightness Tm-doped pump sources.

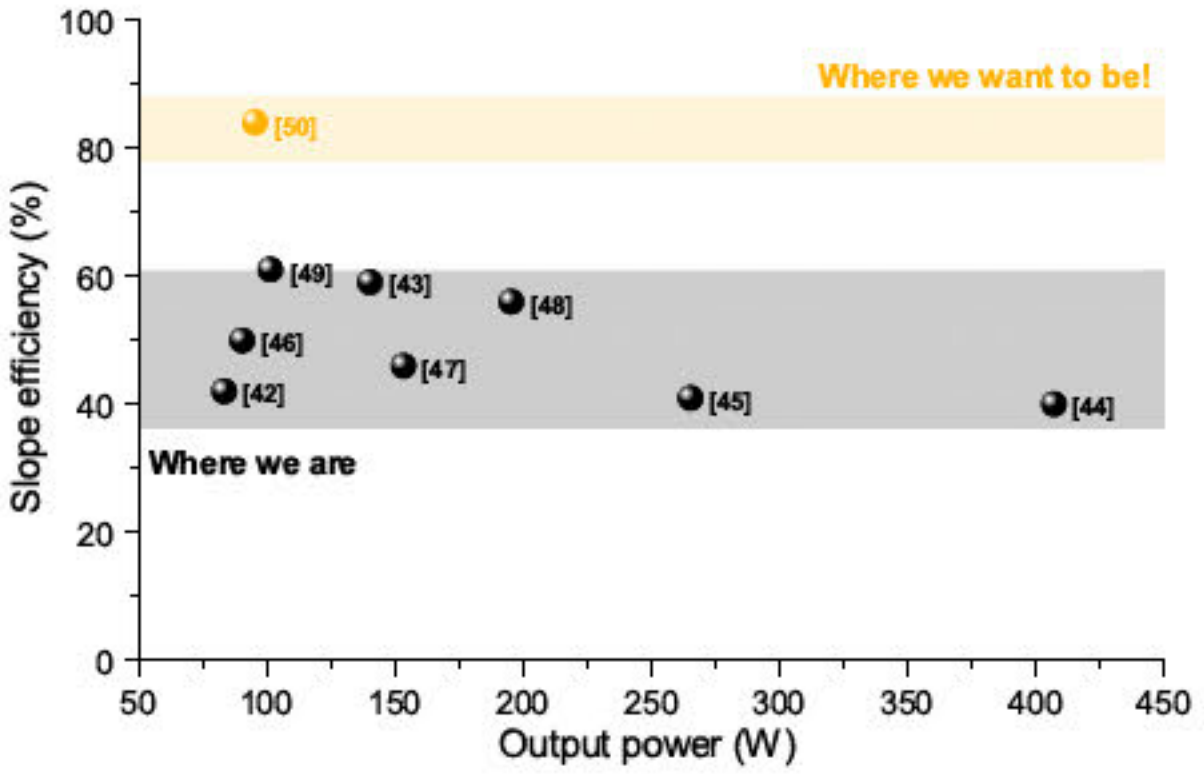

**Figure 8:** Slope efficiency evolution of holmium based fibre lasers and amplifiers with increasing output power [42–50]. The yellow dot represents results for core-pumped fibre laser [50].

Further power scaling approach may be achieved by combining multiple high-power in-band pump sources with a reduced inner-cladding diameter, thereby increasing pump overlap with the doped core and reducing the required active-fibre length. This approach is relevant to both Tm- and Ho-doped systems and can reduce accumulated propagation losses while maintaining the power-handling advantages of cladding pumping. Its practical implementation, however, requires efficient pump combiners and suitable all-glass fibre architectures, together with carefully tailored active fibre [49, 98, 99].

## 5. Nanoparticle doping for highly efficient holmium-doped fibres

Previous work by our research group has shown that highly efficient Ho-doped active fibres require a high Al/Ho ratio, typically of at least 50 [98–100]. Consequently, achieving high-power operation requires sufficiently high concentrations of both $Ho^{3+}$ and $Al_2O_3$. Such compositions are extremely difficult, and in practice may be impossible, to obtain using conventional solution-doping techniques. For this reason, nanoparticle doping was employed in Ho-doped fibres. Advanced chelate doping may provide an alternative route to similarly high dopant concentrations, although at the cost of increased fabrication complexity. In the nanoparticle doping process used here, the practical $Al_2O_3$ concentration is limited to approximately 10 mol. %, corresponding to a maximum $Ho^{3+}$ concentration of about 4000 mol ppm while maintaining the required Al/Ho ratio. This allows high holmium concentrations with a lower fraction of closely interacting $Ho^{3+}$ pairs, thereby minimising pump leakage associated pair-induced quenching (PIQ). Furthermore, $Al_2O_3$ concentrations above 8 mol. % have been shown to provide a lower-phonon-energy local environment for rare-earth ions, which become embedded in alumina-enriched regions, thereby reducing multiphonon relaxation. The reduced non-radiative transition rates prolong the fluorescence lifetime of the $^5I_7 \rightarrow {}^5I_8$ emission, which can in turn be associated with a lower laser threshold.

Complete details of the MCVD preform fabrication and fibre drawing are reported in [98–100]; only the essential steps are summarised here. Fused-silica tubes (F300) with an outer diameter of 18 mm were used as substrates. High-purity $SiCl_4$ and purified oxygen (dew point < -99.9 ° C) served as precursors for $SiO_2$ deposition. The substrate tube was first polished at 2050 ° C, followed by the deposition of two glassy silica buffer layers. A porous core layer was then deposited and subsequently soaked with a doping suspension containing aluminium oxide nanoparticles ($\gamma$-$Al_2O_3$, 99.9 %) and holmium(III) chloride ($HoCl_3{\cdot}6H_2O$, 99.99 %) in absolute ethanol. The doped layer was then dried, sintered under a chlorine atmosphere and collapsed to form the optical preform [98]. The resulting fibres were drawn with a core diameter of 6.5 µm and a cladding diameter of 125 µm (precision ± 1 µm) and coated with a UV-curable acrylate during the drawing process.

Using this approach, a holmium-doped fibre containing 3850 mol ppm of $Ho^{3+}$ and 9.4 mol. % of $Al_2O_3$ was fabricated. The high $Ho^{3+}$ concentration allows the required active-fibre length to be reduced to only several tens of centimetres. This is particularly advantageous in a core-pumped configuration, where the shorter interaction length reduces accumulated propagation losses and can therefore enable a very high conversion efficiency. An all-fibre core-pumped Ho-doped fibre laser resonantly pumped by a Tm-doped fibre laser generated almost 100 W of output power with a slope efficiency of 84 %, approaching the theoretical quantum limit of approximately 91 % [50]. This represents the highest reported slope efficiency for a high-power holmium fibre laser and is more than 25 % higher than values typically achieved in cladding-pumped systems. The work, carried out jointly by the Optoelectronics Research Centre at the University of Southampton and the Institute of Photonics and Electronics of the Czech Academy of Sciences, demonstrates the potential of nanoparticle-assisted high-concentration doping for the development of short, low-loss Ho-doped fibres and highly efficient core-pumped laser architectures.

## 6. Conclusion & Outlook

Recent progress in Tm- and Ho-doped fibre lasers and amplifiers shows that both technologies have reached a level at which further advances increasingly depend on active-fibre design, material optimisation and system integration in addition to the availability of pump power. Although silica fibres have approached their intrinsic attenuation limit at telecommunications wavelengths, propagation losses increase towards and beyond 2 µm due to intrinsic infrared absorption of the silica host material, arising from multiphonon processes. Further optimisation of preform fabrication processes can reduce additional losses associated with rare-earth doping, compositional inhomogeneity and residual

OH contamination. While the intrinsic infrared absorption of silica cannot be eliminated, reducing these additional losses offers a route towards laser efficiencies closer to their theoretical limits. Promising directions include spatially and longitudinally tailored doping, improved rare-earth environments, low-loss all-glass fibre designs, high-power pump combiners and higher-brightness pump sources. Together, these approaches offer routes towards higher efficiency and output power while limiting thermal loading, nonlinear effects and beam-quality degradation.

Similar power-scaling challenges are encountered in other rare-earth-doped fibre lasers, including Yb- and Er and Er/Yb-doped silica fibres and Er fluoride fibres. Although the dominant limiting mechanisms differ, thermal loading remains a common concern, potentially leading to coating damage, thermally induced modal instabilities and beam-quality degradation [101, 102]. Further power scaling therefore requires improved thermal management alongside the mitigation of nonlinear effects and other system-specific limitations.

In Tm-doped systems, kilowatt-level output powers have been demonstrated in both oscillator and MOPA configurations [33, 35, 39]. Further scaling is increasingly linked to improved efficiency and thermal management. Structured radial, multi-element and longitudinally varying $Tm^{3+}$ distributions have shown that the dopant geometry itself can be used to tailor pump absorption and heat generation [37, 74, 70, 85], while optimisation of glass composition can improve the balance between cross-relaxation, propagation loss and thermal loading [60]. Temperature-dependent spectroscopy must also be considered because self-heating can modify gain, ASE and the optimum operating wavelength. Further progress is therefore likely to combine improved glass composition, optimised dopant distributions, advanced fibre geometries and more complete thermo-spectroscopic modelling.

Ho-doped fibre lasers follow a complementary development path. Resonant pumping near 1.9-1.95 μm provides a low quantum defect, while cladding-pumped architectures offer the highest pump-power handling and have enabled substantial output-power scaling [44]. Core-pumped systems, by contrast, can exploit short active-fibre lengths and strong pump-signal overlap to achieve very high efficiency when sufficiently bright Tm-doped pump sources are available [50]. Nanoparticle-assisted fabrication provides a route to the high $Ho^{3+}$ and $Al_2O_3$ concentrations required for such short efficient fibres. Future progress in Ho-doped fibre lasers will therefore remain closely linked both to advances in active-fibre fabrication and to the continued development of high-power and high-brightness Tm-doped pump sources.

## Data availability statement

The data underlying the results of this manuscript are available in the cited references and in the associated repository [78].

## Disclosure

The authors declare that they have no known competing financial interests or personal relationships that could have appeared to influence the work reported in this paper.

## Declaration of generative AI use

During the preparation of this work, the authors used ChatGPT (OpenAI) to improve the English language, grammar, and overall clarity of the manuscript. After using this tool, the authors reviewed and edited the content as needed and take full responsibility for the content of the published paper.

## Acknowledgement

This work was supported by the Czech Science Foundation, No. 26-21543S. This work was co-funded by the European Union and the state budget of the Czech Republic under the project LasApp CZ.02.01.01/00/22_008/0004573.

## Funding

LasApp CZ.02.01.01/00/22_008/0004573; Czech Science Foundation No. 26-21543S.

## CRediT authorship contribution statement

**Richard Švejkar:** Writing – original draft, review & editing, Visualization, Validation, Investigation, Formal analysis. **Pavel Peterka:** Writing – original draft, review & editing, Validation, Supervision, Conceptualization, Methodology, Resources, Project administration, Funding acquisition. **Bára Švejkarová:** Writing – original draft, review & editing, Visualization, Validation, Software, Investigation, Formal analysis. **Michal Kamrádek:** Writing – original draft, review & editing, Investigation. **Petr Vařák:** Writing – original draft, review & editing, Investigation, Formal analysis. **Martin Grábner:** Writing – review & editing, Validation, Software. **Jan Aubrecht:** Writing – review & editing, Investigation, Formal analysis. **Ondřej Podrazký:** Writing – review & editing, Investigation. **Ivo Bartoň:** Investigation. **Ivan Kašík:** Writing – review & editing, Investigation, Supervision, Conceptualization. **Pavel Honzátko:** Investigation, Supervision, Conceptualization, Methodology, Resources, Project administration, Funding acquisition.

## References

[1] Jean-Daniel Colladon. Sur les réflexions d'un rayon de lumière à l'intérieur d'une veine liquide parabolique. *Comptes Rendus Hebdomadaires des Séances de l'Académie des Sciences*, 15:800–802, 1842.

[2] Jacques Babinet. Note sur la transmission de la lumière par des canaux sinueux. *Comptes Rendus Hebdomadaires des Séances de l'Académie des Sciences*, 15:802, 1842.

[3] A. C. S. van Heel. A New Method of Transporting Optical Images without Aberrations. *Nature*, 173:39, 1954. doi:10.1038/173039a0.

[4] T. H. Maiman. Stimulated Optical Radiation in Ruby. *Nature*, 187: 493–494, 1960. doi:10.1038/187493a0.

[5] E. Snitzer. Proposed fiber cavities for optical masers. *Journal of Applied Physics*, 32(1):36–39, 01 1961. ISSN 0021-8979. doi:10.1063/1.1735955.

[6] Charles J. Koester and Elias Snitzer. Amplification in a fiber laser. *Appl. Opt.*, 3(10):1182–1186, Oct 1964. doi:10.1364/AO.3.001182.

[7] J.-C. Simon and E. Spitz. Propagation guidée de lumière cohérente. *Journal de Physique et le Radium*, 24(2):147, 1963. Communication à la Société Française de Physique.

[8] Jeff Hecht. *City of Light: The Story of Fiber Optics*. Oxford University Press, New York, revised and expanded edition, 2004. ISBN 978-0-19-516255-4.

[9] F. N. Steele and R. W. Douglas. Some observations on the absorption of iron in silicate and borate glasses. *Physics and Chemistry of Glasses*, 6(6):246–252, 1965.

[10] K. C. Kao and G. A. Hockham. Dielectric-fibre surface waveguides for optical frequencies. *Proceedings of the Institution of Electrical Engineers*, 113(7):1151–1158, 1966. doi:10.1049/piee.1966.0189.

[11] K. C. Kao and T. W. Davies. Spectrophotometric Studies of Ultra Low Loss Optical Glasses. I. Single Beam Method. *Journal of Physics E: Scientific Instruments*, 1(11):1063–1068, 1968. doi:10.1088/0022-3735/1/11/303.

[12] M. W. Jones and K. C. Kao. Spectrophotometric Studies of Ultra Low Loss Optical Glasses. II. Double Beam Method. *Journal of Physics E: Scientific Instruments*, 2(4):331–335, 1969.

[13] C. R. Wright and K. C. Kao. Spectrophotometric Studies of Ultra Low Loss Optical Glasses. III. Ellipsometric Determination of Surface Reflectances. *Journal of Physics E: Scientific Instruments*, 2:579–583, 1969.

[14] X. Steve Yao and Xiaojun Chen. *Polarization Measurement and Control in Optical Fiber Communication and Sensor Systems*. John Wiley & Sons, 2022. ISBN 9781119758471. doi:10.1002/9781119758488.

[15] F. P. Kapron, D. B. Keck, and R. D. Maurer. Radiation Losses in Glass Optical Waveguides. In *Proceedings of the Conference on Trunk Telecommunications by Guided Waves*, pages 148–153, London, UK, September 1970. Institution of Electrical Engineers.

[16] E. Snitzer. Optical Maser Action of $Nd^{3+}$ in a Barium Crown Glass. *Physical Review Letters*, 7(12):444–446, 1961. doi:10.1103/PhysRevLett.7.444.

[17] S. B. Poole, D. N. Payne, and M. E. Fermann. Fabrication of Low-Loss Optical Fibres Containing Rare-Earth Ions. *Electronics Letters*, 21(17):737–738, 1985. doi:10.1049/el:19850520.

[18] I.P. Alcock, A.C. Tropper, A.I. Ferguson, and D.C. Hanna. Q-switched operation of a neodymium-doped monomode fibre laser. *Electronics Letters*, 22:84–85, 1986. doi:10.1049/el:19860057.

[19] D.C. Hanna, I.M. Jauncey, R.M. Percival, I.R. Perry, R.G. Smart, P.J. Suni, J.E. Townsend, and A.C. Tropper. Continuous-wave oscillation of a monomode thulium-doped fibre laser. *Electronics Letters*, 24:1222–1223, 1988. doi:10.1049/el:19880831.

[20] D.C. Hanna, I.R. Perry, J.R. Lincoln, and J.E. Townsend. A 1-Watt thulium-doped cw fibre laser operating at 2 μm. *Optics Communications*, 80(1):52–56, 1990. ISSN 0030-4018. doi:10.1016/0030-4018(90)90505-N.

[21] Stuart D. Jackson and Terence A. King. High-power diode-cladding-pumped Tm-doped silica fiber laser. *Opt. Lett.*, 23(18):1462–1464, Sep 1998. doi:10.1364/OL.23.001462.

[22] R.A. Hayward, W.A. Clarkson, P.W. Turner, J. Nilsson, A.B. Grudinin, and D.C. Hanna. Efficient cladding-pumped Tm-doped silica fibre laser with high power singlemode output at 2 μm. *Electronics Letters*, 36:711–712, 2000. doi:10.1049/el:20000577.

[23] Gavin P. Frith and David G. Lancaster. Power scalable and efficient 790-nm pumped $Tm^{3+}$-doped fiber lasers. In Andrew J. W. Brown, Johan Nilsson, Donald J. Harter, and Andreas Tünnermann, editors, *Fiber Lasers III: Technology, Systems, and Applications*, volume 6102, page 610208. International Society for Optics and Photonics, SPIE, 2006. doi:10.1117/12.660932.

[24] Christophe Louot, Félix Sanson, Arnaud Motard, Thierry Ibach, Inka Manek-Hönninger, Antoine Berrou, Nicolas Dalloz, Thierry Robin, Benoit Cadier, and Anne Hildenbrand-Dhollande. Emission Wavelength Limits of a Continuous-Wave Thulium-Doped Fiber Laser Source Operating at 1.94 μm, 2.09 μm or 2.12 μm. *Photonics*, 11(3), 2024. ISSN 2304-6732. doi:10.3390/photonics11030246.

[25] Peter F. Moulton, Glen A. Rines, Evgueni V. Slobodtchikov, Kevin F. Wall, Gavin Frith, Bryce Samson, and Adrian L. G. Carter. Tm-Doped Fiber Lasers: Fundamentals and Power Scaling. *IEEE Journal of Selected Topics in Quantum Electronics*, 15(1):85–92, 2009. doi:10.1109/JSTQE.2008.2010719.

[26] Gregory D. Goodno, Lewis D. Book, and Joshua E. Rothenberg. Low-phase-noise, single-frequency, single-mode 608 W thulium fiber amplifier. *Opt. Lett.*, 34(8):1204–1206, Apr 2009. doi:10.1364/OL.34.001204.

[27] Thomas Ehrenreich, Ryan Leveille, Imtiaz Majid, Kanishka Tankala, Glen Rines, and Peter Moulton. 1-kW, All-Glass Tm:fiber Laser. volume 7580. SPIE, 2010. URL https://www.yumpu.com/en/document/view/34901571/pw-2010-1-kw-all-glass-tmfiber-laser-q-peak-inc. unpublished by SPIE.

[28] Benjamin R. Johnson, Daniel Creeden, Julia Limongelli, Herman Pretorius, Jon Blanchard, and Scott D. Setzler. Comparison of high power large mode area and single mode 1908nm Tm-doped fiber lasers. In John Ballato, editor, *Fiber Lasers XIII: Technology, Systems, and Applications*, volume 9728, page 972810. International Society for Optics and Photonics, SPIE, 2016. doi:10.1117/12.2209677.

[29] Till Walbaum, Matthias Heinzig, Thomas Schreiber, Ramona Eberhardt, and Andreas Tünnermann. Monolithic thulium fiber laser with 567 W output power at 1970 nm. *Opt. Lett.*, 41(11):2632–2635, Jun 2016. doi:10.1364/OL.41.002632.

[30] C. Gaida, M. Gebhardt, T. Heuermann, F. Stutzki, C. Jauregui, and J. Limpert. Ultrafast thulium fiber laser system emitting more than 1 kW of average power. *Opt. Lett.*, 43(23):5853–5856, Dec 2018. doi:10.1364/OL.43.005853.

[31] Liu Yin-Zi, Xing Ying-Bing, Liao Lei, Wang Yi-Bo, Peng Jing-Gang, Li Hai-Qing, Dai Neng-Li, and Li Jin-Yan. 530 w all-fiber continuous-wave tm-doped fiber laser. *Acta Physica Sinica*, 69(18):184209–1–184209–7, 2020. ISSN 1000-3290. doi:10.7498/aps.69.20200466.

[32] Brian M. Anderson, Joel Solomon, and Angel Flores. 1.1 kW, beam-combinable thulium doped all-fiber amplifier. In Michalis N. Zervas, editor, *Fiber Lasers XVIII: Technology and Systems*, volume 11665, page 116650B. International Society for Optics and Photonics, SPIE, 2021. doi:10.1117/12.2576209.

[33] Chuanyong Ren, Yiqi Shen, Yuhao Zheng, Yuji Mao, Fei Wang, Deyuan Shen, and Heyuan Zhu. Widely-tunable all-fiber tm doped mopa with > 1 kW of output power. *Opt. Express*, 31(14):22733–22739, Jul 2023. doi:10.1364/OE.494015.

[34] Maria Michalska, Pavel Honzatko, Pawel Grzes, Michal Kamradek, Ondrej Podrazky, Ivan Kasik, and Jacek Swiderski. Thulium-Doped 1940- and 2034-nm Fiber Amplifiers: Towards Highly Efficient, High-Power All-Fiber Laser Systems. *J. Lightwave Technol.*, 42(1):339–346, Jan 2024. doi:10.1109/JLT.2023.3301397.

[35] Clément Romano, Dieter Panitzek, Dominik Lorenz, Patrick Forster, Marc Eichhorn, and Christelle Kieleck. High-Power Thulium-Doped Fiber MOPA Emitting at 2036 nm. *Journal of Lightwave Technology*, 42(1):394–398, 2024. doi:10.1109/JLT.2023.3310121.

[36] Hongyu Wang, Qilai Zhao, Hang Liu, Yuxin Sun, Jialong Li, Junjie Zheng, Ye Yuan, Qianwen Zhang, Changsheng Yang, Yujun Feng, Yinhong Sun, Zhongmin Yang, and Shanhui Xu. Kilowatt-Level High-Efficiency Narrow-Linewidth All-Fiber $Tm^{3+}$-Doped Laser. *Photonics*, 11(9), 2024. ISSN 2304-6732. doi:10.3390/photonics11090877.

[37] Richard Švejkar, Martin P. Buckthorpe, Bára Švejkarová, Peter C. Shardlow, and W. Andrew Clarkson. Monolithic highly efficient 272 W nested-ring structured-core thulium fiber laser at 1940 nm. *Opt. Lett.*, 50(18):5614–5617, Sep 2025. doi:10.1364/OL.566019.

[38] Félix Sanson, Christophe Louot, Achille Bogas-Droy, Stefano Bigotta, Caterina Clemente, Nicolas Dalloz, Arnaud Motard, Inka Manek-Hönninger, and Anne Hildenbrand-Dhollande. 680 W thulium-doped all-fiber single-oscillator laser emitting at 2036-nm in continuous-wave operation. *Opt. Continuum*, 4(7):1426–1432, Jul 2025. doi:10.1364/OPTCON.566656.

[39] Jun Ho Lee, Gaye Park, Jaesun Kim, Ji Won Kim, and Hoon Jeong. All-fiberized continuous-wave thulium-doped fiber laser with > 1 kw output power at 1940 nm. *Optics Communications*, 615:133314, 2026. ISSN 0030-4018. doi:10.1016/j.optcom.2026.133314.

[40] Kyunghwan Oh, T. F. Morse, A. Kilian, L. Reinhart, and P. M. Weber. Continuous-wave oscillation of thulium-sensitized holmium-doped silica fiber laser. *Opt. Lett.*, 19(4):278–280, Feb 1994. doi:10.1364/OL.19.000278.

[41] Yahua Li, Yucheng Zhao, B. Ashton, S. Jackson, and S. Fleming. Highly efficient and wavelength-tunable Holmium-doped silica fibre lasers. In *2005 31st European Conference on Optical Communication, ECOC 2005*, volume 3, pages 679–680 vol.3, 2005. doi:10.1049/cp:20050668.

[42] Stuart D. Jackson, Alex Sabella, Alex Hemming, Shayne Bennetts, and David G. Lancaster. High-power 83 W holmium-doped silica fiber laser operating with high beam quality. *Opt. Lett.*, 32(3):241–243, Feb 2007. doi:10.1364/OL.32.000241.

[43] Alexander Hemming, Shayne Bennetts, Nikita Simakov, Alan Davidson, John Haub, and Adrian Carter. High power operation of cladding pumped holmium-doped silica fibre lasers. *Opt. Express*, 21(4):4560–4566, Feb 2013. doi:10.1364/OE.21.004560.

[44] Alexander Hemming, Nikita Simakov, John Haub, and Adrian Carter. A review of recent progress in holmium-doped silica fibre sources. *Optical Fiber Technology*, 20(6):621–630, 2014. ISSN 1068-5200. doi:10.1016/j.yofte.2014.08.010.

[45] Alexander Hemming, Nikita Simakov, Alan Davidson, Michael Oermann, Len Corena, Dmitrii Stepanov, Neil Carmody, John Haub, Robert Swain, and Adrian Carter. Development of high-power holmium-doped fibre amplifiers. In Siddharth Ramachandran, editor, *Fiber Lasers XI: Technology, Systems, and Applications*, volume 8961, page 89611A. International Society for Optics and Photonics, SPIE, 2014. doi:10.1117/12.2042963.

[46] Julien Le Gouët, François Gustave, Pierre Bourdon, Thierry Robin, Arnaud Laurent, and Benoit Cadier. Realization and simulation of high-power holmium doped fiber lasers for long-range transmission. *Opt. Express*, 28(15):22307–22320, Jul 2020. doi:10.1364/OE.394011.

[47] Arnaud Motard, Nicolas Dalloz, Christophe Louot, Thierry Robin, Benoit Cadier, Inka Manek-Hönninger, and Anne Dhollande-Hildenbrand. 150-W continuous wave laser emission at 2.09 μm from a $Tm^{3+}$, $Ho^{3+}$-codoped single-oscillator monolithic fiber laser. In *9TH EPS-QEOD EUROPHOTON VIRTUAL CONFERENCE*, virtual, Czech Republic, August 2020. URL https://hal.science/hal-03413887.

[48] Arnaud Motard, Christophe Louot, Thierry Robin, Benoit Cadier, Inka Manek-Hönninger, Nicolas Dalloz, and Anne Hildenbrand-Dhollande. Diffraction limited 195-W continuous wave laser emission at 2.09 μm from a $Tm^{3+}$,$Ho^{3+}$-codoped single-oscillator monolithic fiber laser. *Opt. Express*, 29(5):6599–6607, Mar 2021. doi:10.1364/OE.416443.

[49] Nicolas Dalloz, Stefano Bigotta, Thierry Ibach, Christophe Louot, Thierry Robin, and Anne Hildenbrand-Dhollande. Triple-Clad Fiber Combiner for Holmium-Doped Fiber Lasers Clad-Pumping. *Photonics*, 12(7), 2025. ISSN 2304-6732. doi:10.3390/photonics12070659.

[50] Richard Švejkar, Bára Švejkarová, Michal Kamrádek, Jan Aubrecht, Pavel Peterka, and W. Andrew Clarkson. All-fiber core-pumped ho-doped laser with 95 W output power and high efficiency. *Opt. Express*, 34(6):10971–10977, Mar 2026. doi:10.1364/OE.589906.

[51] Nathaniel M. Fried and Keith E. Murray. High-power thulium fiber laser ablation of urinary tissues at 1.94 μm. *Journal of Endourology*, 19(1):25–31, 2005. doi:10.1089/end.2005.19.25.

[52] Ilya Mingareev, Fabian Weirauch, Alexander Olowinsky, Lawrence Shah, Pankaj Kadwani, and Martin Richardson. Welding of polymers using a 2 μm thulium fiber laser. *Optics & Laser Technology*, 44(7):2095–2099, 2012. doi:10.1016/j.optlastec.2012.03.020.

[53] Mengmeng Tao, Bo Tao, Zhiyun Hu, Guobin Feng, Xisheng Ye, and Jun Zhao. Development of a 2 μm tm-doped fiber laser for hyperspectral absorption spectroscopy applications. *Optics Express*, 25(26):32386–32394, 2017. doi:10.1364/OE.25.032386.

[54] Daniel Creeden, Peter A. Ketteridge, Peter A. Budni, Scott D. Setzler, York E. Young, John C. McCarthy, Kevin Zawilski, Peter G. Schunemann, Thomas M. Pollak, Evan P. Chicklis, and Min Jiang. Mid-infrared $ZnGeP_2$ parametric oscillator directly pumped by a pulsed 2 μm Tm-doped fiber laser. *Optics Letters*, 33(4):315–317, 2008. doi:10.1364/OL.33.000315.

[55] Qing Wang, Jihong Geng, and Shibin Jiang. 2-μm fiber laser sources for sensing. *Optical Engineering*, 53(6):061609, 2014. doi:10.1117/1.OE.53.6.061609.

[56] C. W. Rudy, M. J. F. Digonnet, and R. L. Byer. Advances in 2-μm Tm-doped mode-locked fiber lasers. *Optical Fiber Technology*, 20 (6):642–649, 2014. doi:10.1016/j.yofte.2014.06.005.

[57] E. Snitzer, H. Po, F. Hakimi, R. Tumminelli, and B. C. McCollum. Double-Clad, Offset Core Nd Fiber Laser. In *Optical Fiber Sensors*, volume 2 of *OSA Technical Digest Series*, page PD5. Optical Society of America, 1988. doi:10.1364/OFS.1988.PD5.

[58] Stuart D. Jackson and Simon Mossman. Efficiency dependence on the $Tm^{3+}$ and $Al^{3+}$ concentrations for $Tm^{3+}$-doped silica double-clad fiber lasers. *Applied Optics*, 42(15):2702–2707, 2003. doi:10.1364/AO.42.002702.

[59] Alex Sincore, Joshua D. Bradford, Justin Cook, Lawrence Shah, and Martin C. Richardson. High Average Power Thulium-Doped Silica Fiber Lasers: Review of Systems and Concepts. *IEEE Journal of Selected Topics in Quantum Electronics*, 24(3):0901808, 2018. doi:10.1109/JSTQE.2017.2775964.

[60] Richard Švejkar, Nilotpal Choudhury, Tom G. Harris, Martin P. Buckthorpe, Jayanta K. Sahu, and W. Andrew Clarkson. Dependence of quantum efficiency and core propagation loss on doping concentration in Tm-doped silica fibers. *Opt. Express*, 33(23): 48807–48814, Nov 2025. doi:10.1364/OE.578867.

[61] Mark D. Burns, Peter C. Shardlow, and W. Andrew Clarkson. Widely-Tunable Operation of a Thulium Doped Fibre Laser Between 1654 nm and 2025 nm. In *2021 Conference on Lasers and Electro-Optics Europe and European Quantum Electronics Conference (CLEO/Europe-EQEC)*, pages 1–1, 2021. doi:10.1109/CLEO/Europe-EQEC52157.2021.9542539.

[62] Jan Aubrecht, Jan Pokorný, Bára Švejkarová, Michal Kamrádek, and Pavel Peterka. Broadband thulium fiber amplifier for spectral region located beyond the l-band. *Optics Express*, 32(10):17932–17941, 2024. doi:10.1364/OE.522088.

[63] Jan Šulc, Michal Němec, Jan Kratochvíl, and Helena Jelínková. Tm-fiber laser resonantly pumped by 1.7 μm laser diode. In Clémence Jollivet, editor, *Fiber Lasers XXI: Technology and Systems*, volume 12865, page 128651A. International Society for Optics and Photonics, SPIE, 2024. doi:10.1117/12.3000072.

[64] Mathias Lenski, Felix Wanitschke, Florian Bontke, Mehran Bahri, Philipp Gierschke, Christian Gaida, Tobias Heuermann, César Jáuregui, and Jens Limpert. Power and energy scaling of tm-doped fiber lasers enabled by in-band pumping. *Opt. Express*, 34(19): 35811–35820, Sep 2026. doi:10.1364/OE.601176.

[65] Jan Pokorný, Jan Aubrecht, Michal Kamrádek, Bára Švejkarová, Petr Vařák, Martin Grábner, and Pavel Peterka. Depressed-cladding thulium-doped fiber for applications below 1800 nm. *Opt. Express*, 32(10):17966–17976, May 2024. doi:10.1364/OE.523168.

[66] Junxiang Zhang, Shijie Fu, Quan Sheng, Lu Zhang, Wei Shi, and Jianquan Yao. Recent progress on power scaling and single-frequency operation of 1.7-μm thulium-doped fiber lasers. *Optics & Laser Technology*, 158:108882, 2023. ISSN 0030-3992. doi:10.1016/j.optlastec.2022.108882.

[67] Kentin Poncelet, Thomas Dubé, Coline Lavit, Ruizhe Gu, Saïd Hamdi, Roopa Prakash, Nicholas Traynor, Adèle Hilico, Germain Guiraud, and Giorgio Santarelli. 30 W-class tunable single-frequency thulium-doped fiber amplifier at 1730–1762 nm with low intensity noise. *Opt. Lett.*, 51(9):2564–2567, May 2026. doi:10.1364/OL.597210.

[68] Nikita Simakov, Alexander V. Hemming, Adrian Carter, Kevin Farley, Alan Davidson, Neil Carmody, Mark Hughes, Jae M. O. Daniel, Len Corena, Dmitrii Stepanov, and John Haub. Design and experimental demonstration of a large pedestal thulium-doped fibre. *Opt. Express*, 23(3):3126–3133, Feb 2015. doi:10.1364/OE.23.003126.

[69] H. Lopez Aviles, B. Faugas, J. Bradford, A. Agliati, S. Morris, and C. Jollivet. Single-mode operation of a pedestal-free 25 μm core double-clad Tm-doped fiber with > 50 % efficiency for 2 μm applications. In Pavel Peterka, Kyriacos Kalli, John Ballato, and Alexis Mendez, editors, *Eighth International Workshop on Specialty Optical Fibers and Their Applications (WSOF 2025)*, volume 13522, page 1352214. International Society for Optics and Photonics, SPIE, 2025. doi:10.1117/12.3056562.

[70] Matthew J. Barber, Peter C. Shardlow, Pranabesh Barua, Jayanta K. Sahu, and W. Andrew Clarkson. Nested-ring doping for highly efficient 1907 nm short-wavelength cladding-pumped thulium fiber lasers. *Optics Letters*, 45(19):5542–5545, 2020. doi:10.1364/OL.401228.

[71] Stefan Kuhn, Bruno Poletto Rodrigues, Dintu Mathew, Sigrun Hein, Johannes Nold, Nicoletta Haarlammert, and Thomas Schreiber. The effect of fluorine codoping on the fluorescence properties of tm-doped fiber laser preforms. In *Optica Laser Congress Congress 2025*, page JTu2A.41. Optica Publishing Group, 2025. doi:10.1364/ASSL.2025.JTu2A.41.

[72] Petr Vařák, Martin Leich, Michal Kamrádek, Jan Aubrecht, Ondřej Podrazký, Ivo Bartoň, Bára Švejkarová, Alena Michalcová, Katrin Wondraczek, Matthias Jäger, Ivan Kašík, Pavel Peterka, and Pavel Honzátko. Nanoparticle doping and molten-core methods towards highly thulium-doped silica fibers for 0.79 μm pumped 2 μm fiber lasers – A fluorescence lifetime study. *Journal of Luminescence*, 275:120835, 2024. doi:10.1016/j.jlumin.2024.120835.

[73] Michal Kamrádek, Jan Aubrecht, Petr Vařák, Jakub Cajzl, Václav Kubeček, Pavel Honzátko, Ivan Kašík, and Pavel Peterka. Energy transfer coefficients in thulium-doped silica fibers. *Opt. Mater. Express*, 11(6):1805–1814, Jun 2021. doi:10.1364/OME.427456.

[74] Ivo Bartoň, Jan Aubrecht, Bára Švejkarová, Jan Pokorný, Michal Kamrádek, Ondřej Podrazký, Petr Vařák, Ivan Kašík, Martin Grábner, and Pavel Peterka. Pedestal-free 7-rod-core Tm-doped fiber for fiber lasers. *Optics Express*, 34(9):16838–16847, 2026. doi:10.1364/OE.591864.

[75] Martin Grábner, Pavel Peterka, and Pavel Honzátko. Formula for temperature distribution in multi-layer optical fibres for high-power fibre lasers. *Opto-Electronics Review*, 29(4):126–132, 2021. doi:10.24425/opelre.2021.139482.

[76] Bára Jiříčková, Richard Švejkar, Martin Grábner, Cesar Jauregui, Jan Aubrecht, Ondřej Schreiber, and Pavel Peterka. Temperature-dependent emission cross-section spectra at 1.8 μm of the thulium-doped fibers cooled down to cryogenic temperatures. In *Conference on Lasers and Electro-OpticsEurope (CLEOEurope 2023) and European Quantum Electronics Conference (EQEC 2023)*. Optica Publishing Group, 2023. URL `https://opg.optica.org/abstract.cfm?URI=CLEO_Europe-2023-cj_10_6`.

[77] Bára Jiříčková, Martin Grábner, Cesar Jauregui, Jan Aubrecht, Ondřej Schreiber, and Pavel Peterka. Temperature-dependent cross section spectra for thulium-doped fiber lasers. *Optics Letters*, 48(3): 811–814, 2023. doi:10.1364/OL.479313.

[78] Richard Švejkar, Pavel Peterka, Bára Švejkarová, Michal Kamrádek, Petr Vařák, Martin Grábner, Jan Aubrecht, Ondřej Podrazký, Ivo Bartoň, Ivan Kašík, and Pavel Honzátko. Supplementary data for review paper called thulium- and holmium-doped silica fibres for 2 μm fibre lasers, October 2026. URL `https://doi.org/10.5281/zenodo.23759388`.

[79] G. Pan, N. Yu, B. Meehan, T. W. Hawkins, J. Ballato, and P. D. Dragic. Thermo-optic coefficient of b2o3 and geo2 co-doped silica fibers. *Opt. Mater. Express*, 10(7):1509–1521, Jul 2020. doi:10.1364/OME.397215.

[80] Pavel Peterka, Bára Jiříčková, Ondřej Schreiber, and Martin Grábner. Thulium Cross Sections Temperature Dependence and Its Effect on Fiber Laser Operation. In *2023 23rd International Conference on Transparent Optical Networks (ICTON)*, pages 1–4, Bucharest, Romania, 2023. doi:10.1109/ICTON59386.2023.10207214.

[81] Martin Grábner, Bára Švejkarová, Jan Aubrecht, and Pavel Peterka. Analytical Model of Thulium-Doped Fiber Laser Pumped by Two-for-One Process. *Journal of Lightwave Technology*, 42(8):2938–2944, 2024. doi:10.1109/JLT.2023.3345001.

[82] Martin Grábner, Bára Švejkarová, Jan Aubrecht, Jan Pokorný, Pavel Honzátko, and Pavel Peterka. Rise of amplified spontaneous emission in high-power thulium-doped fiber lasers and amplifiers due to self-heating. *Optics & Laser Technology*, 180:111428, 2025. doi:10.1016/j.optlastec.2024.111428.

[83] I. Kasik, I. Barton, M. Kamradek, O. Podrazky, J. Aubrecht, P. Varak, P. Peterka, P. Honzatko, D. Pysz, M. Franczyk, and R. Buczynski. Doped and structured silica optical fibres for fibre laser sources. *Optics Communications*, 577:131437, 2025. ISSN 0030-4018. doi:10.1016/j.optcom.2024.131437.

[84] D. Pysz, I. Kujawa, R. Stępień, M. Klimczak, A. Filipkowski, M. Franczyk, L. Kociszewski, J. Buźniak, K. Haraśny, and R. Buczyński. Stack and draw fabrication of soft glass microstructured fiber optics. *Bulletin of the Polish Academy of Sciences Technical Sciences*, 62(No 4):667–682, 2014. doi:10.2478/bpasts-2014-0073.

[85] Bára Švejkarová, Martin Grábner, Jan Aubrecht, Richard Švejkar, Jan Pokorný, Michal Kamrádek, Ondřej Podrazký, Ivan Kašík, Pavel Honzátko, W. Andrew Clarkson, and Pavel Peterka. Thulium fiber lasers with longitudinally modified concentration. *High Power Laser Science and Engineering*, 13:e71, 2025. doi:10.1017/hpl.2025.10051.

[86] Pavel Peterka, Ivan Kašík, Ondřej Podrazký, Michal Kamrádek, and Pavel Honzátko. Active fibers for 2 μ m fiber lasers. In *Laser Congress 2023 (ASSL, LAC)*, page ATh4A.1. Optica Publishing Group, 2023. doi:10.1364/ASSL.2023.ATh4A.1.

[87] Bára Švejkarová, Jan Aubrecht, Ivo Bartoň, Richard Švejkar, Michal Kamrádek, Petr Vařák, Ondřej Podrazký, Zuzana Záborská, and Pavel Peterka. Structured-core thulium-doped fiber for efficient high-power laser operation at 1907 nm. In *Optica Laser Congress Congress 2026*, page HTh2C.6. Optica Publishing Group, 2026. to be publish.

[88] Pavel Peterka, Jan Aubrecht, Dariusz Pysz, Marcin Franczyk, Ondřej Schreiber, Michal Kamrádek, Ivan Kašík, and Ryszard Buczyński. Development of pedestal-free large mode area fibers with $Tm^{3+}$ doped silica nanostructured core. *Optics Express*, 31(26):43004–43016, 2023. doi:10.1364/OE.503047.

[89] Przemysław Gołębiewski, Paweł Wiencław, Jarosław Cimek, Paweł Socha, Dariusz Pysz, Adam Filipkowski, Grzegorz Stępniewski, Olga Czerwińska, Ireneusz Kujawa, Ryszard Stępień, Rafał Kasztelanic, Andrzej Burgs, and Ryszard Buczyński. 3D soft glass printing of preforms for microstructured optical fibers. *Additive Manufacturing*, 79:103899, 2024. ISSN 2214-8604. doi:10.1016/j.addma.2023.103899.

[90] Lars G. Holmen and Helge Fonnum. Holmium-doped fiber amplifier for pumping a $ZnGeP_2$ optical parametric oscillator. *Opt. Express*, 29(6):8477–8489, Mar 2021. doi:10.1364/OE.419939.

[91] Jan Pokorný, Bára Švejkarová, Jan Aubrecht, Michal Kamrádek, Ivo Bartoň, Ivan Kašík, Pavel Honzátko, and Pavel Peterka. Holmium-doped silica fibers combining high doping and high efficiency. *Optics Express*, 33(7):14843–14849, 2025. doi:10.1364/OE.554762.

[92] Guillaume G. Vienne, Julie E. Caplen, Liang Dong, John D. Minelly, Johan Nilsson, and David N. Payne. Fabrication and Characterization of $Yb^{3+}$:$Er^{3+}$ Phosphosilicate Fibers for Lasers. *Journal of Lightwave Technology*, 16(11):1990–2001, 1998.

doi:10.1109/50.730360.
[93] Lars G. Holmen, Peter C. Shardlow, Pranabesh Barua, Jayanta K. Sahu, Nikita Simakov, Alexander Hemming, and W. Andrew Clarkson. Tunable holmium-doped fiber laser with multiwatt operation from 2025 nm to 2200 nm. *Opt. Lett.*, 44(17):4131–4134, Sep 2019. doi:10.1364/OL.44.004131.
[94] Colin C. Baker, E. Joseph Friebele, Ashley A. Burdett, L. Brandon Shaw, Steven R. Bowman, Woohong Kim, Jasbinder S. Sanghera, John M. Ballato, Courtney Kucera, Amber Vargas, Alexander V. Hemming, Nikita Simikov, and John Haub. Recent advances in holmium doped fibers for high-energy lasers (Conference Presentation). In Mark Dubinskiy and Timothy C. Newell, editors, *Laser Technology for Defense and Security XIV*, volume 10637, page 1063704. International Society for Optics and Photonics, SPIE, 2018. doi:10.1117/12.2310134.
[95] Jiachen Wang, Dong-il Yeom, Nikita Simakov, Alexander Hemming, Adrian Carter, Sang Bae Lee, and Kwanil Lee. Numerical Modeling of in-Band Pumped Ho-Doped Silica Fiber Lasers. *Journal of Lightwave Technology*, 36(24):5863–5880, 2018. doi:10.1109/JLT.2018.2877817.
[96] Martin Grábner, Kanagaraj Nithyanandan, Pavel Peterka, Pavel Koška, Ali. A. Jasim, and Pavel Honzátko. Simulations of Pump Absorption in Tandem-Pumped Octagon Double-Clad Fibers. *IEEE Photonics Journal*, 13(2):1–14, 2021. doi:10.1109/JPHOT.2021.3060857.
[97] Daniel Creeden, Benjamin R. Johnson, Glen A. Rines, and Scott D. Setzler. High power resonant pumping of tm-doped fiber amplifiers in core- and cladding-pumped configurations. *Opt. Express*, 22(23): 29067–29080, Nov 2014. doi:10.1364/OE.22.029067.
[98] Michal Kamrádek, Ivan Kašík, Jan Aubrecht, Petr Vařák, Ondřej Podrazký, Ivo Bartoň, Jan Pokorný, Pavel Peterka, and Pavel Honzátko. Nanoparticle doping as a way to enhance holmium fiber lasers efficiency. *Optics Communications*, 575:131290, 2025. doi:10.1016/j.optcom.2024.131290.
[99] Michal Kamrádek, R. Sajzew, Jan Aubrecht, F. Lindner, Petr Vařák, Ondřej Podrazký, Ivo Bartoň, J. Bierlich, J. Dellith, Ivan Kašík, K. Wondraczek, Pavel Peterka, and Pavel Honzátko. The role of composition and preparation method on laser performance of holmium-doped silica fibers. *Optical Materials: X*, 30:100440, 2026. doi:10.1016/j.omx.2026.100440.
[100] Michal Kamrádek. 7 - Nanoparticles-doped silica-glass-based optical fibers: fabrication and application. In Mário F.S. Ferreira and Mukul Chandra Paul, editors, *Specialty Optical Fibers*, Woodhead Publishing Series in Electronic and Optical Materials, pages 159–181. Woodhead Publishing, 2024. ISBN 978-0-443-18495-6. doi:10.1016/B978-0-443-18495-6.00007-X.
[101] Michalis N. Zervas and Christophe A. Codemard. High Power Fiber Lasers: A Review. *IEEE Journal of Selected Topics in Quantum Electronics*, 20(5):219–241, 2014. doi:10.1109/JSTQE.2014.2321279.
[102] Michalis N. Zervas. Transverse mode instability, thermal lensing and power scaling in $Yb^{3+}$-doped high-power fiber amplifiers. *Opt. Express*, 27(13):19019–19041, Jun 2019. doi:10.1364/OE.27.019019.